# Effect of THz-radiation on Behavior of Male Mice

N. P. Bondar, I. L. Kovalenko, D. F. Avgustinovich, *A.G.Khamoyan, N. N. Kudryavtseva

Institute of Cytology and Genetics SD RAS,  *Institute of Laser Physics SD RAS, Novosibirsk, Russia

Effect of terahertz radiation (3.6 THz, 81.5 mkm, 15 mV) on some behavioral patterns of intact mice has been investigated. In home cage mice demonstrated avoidance of laser ray and enhanced replacement activity in free behavior. Animals irradiated during 30 minutes manifested an increased level of anxiety, which was evaluated in the plus maze test on the day following the radiation.

---

# ВЛИЯНИЕ ЭЛЕКТРОМАГНИТНОГО ИЗЛУЧЕНИЯ ТЕРАГЕРЦЕВОГО ДИАПАЗОНА НА ПОВЕДЕНИЕ САМЦОВ МЫШЕЙ

Н.П.Бондарь, И.Л. Коваленко, Д.Ф. Августинович, *А. Г. Хамоян., Н.Н. Кудрявцева.

Институт цитологии и генетики СО РАН,  *Институт лазерной физики СО РАН, Новосибирск, natnik@bionet.nsc.ru

Изучено влияние электромагнитного излучения терагерцевого диапазона (3,6 ТГц, 81,5 мкм, 15 мВт) на некоторые формы поведения интактных мышей. В домашней клетке мыши демонстрировали избегание зоны действия луча лазера и повышенную смещенную активность в свободном поведении. Облучение в течение 30 мин повышало у животных уровень тревожности, оцениваемый в крестообразном лабиринте на следующий день после облучения.

**Ключевые слова**: ТГц-волны, тревожность, крестообразный лабиринт, мыши, поведение.

---

Диапазон терагерцевого излучения (ТГц-волны) лежит на границе между радиоволнами (микроволновый интервал) и светом (инфракрасный интервал). В терминах частот электромагнитных колебаний этот диапазон находится в пределах от 0,1 ТГц до 10 ТГц (T-rays). Полагают, что эти волны способны проникать сквозь некоторые твердые вещества, не обладают вредным ионизирующим воздействием на биологические объекты и именно с этими свойствами ТГц-волн связывают будущие успехи в области медицины, работы служб безопасности, телекоммуникаций, инфотехнологий и защиты окружающей среды [1, 2, 7, 8, 9, 11]. Однако, из-за очень малой мощности имеющихся источников терагерцевого излучения по сравнению с инфракрасным и микроволновым спектром, ТГц-волны намного труднее обнаруживать и анализировать. Поэтому до настоящего времени эффект воздействия ТГц-волн на биологические объекты разного уровня сложности мало изучен. Целью данного исследования было ответить на принципиальный вопрос: чувствуют ли животные терагерцевое излучение? В качестве возможных параметров, свидетельствующих о влиянии терагерцевого излучения на живой организм, были исследованы некоторые формы поведения самцов мышей, находящихся в поле действия луча лазера, работающего в терагерцевой области, а также был изучен отставленный эффект облучения ТГц-волнами на последующее поведение мышей. Работы проводили с использованием лазера, разработанного (патент № 2143162) и исследованного [3, 10] на базе Института лазерной физики СО РАН, Новосибирск. Частота излучения лазера равнялась 3,6 ТГц при длине волны 81,5 мкм и мощности ≈ 15 мВт.

## МАТЕРИАЛЫ И МЕТОДЫ

Эксперименты проводили на половозрелых самцах мышей линии C57BL/6J в возрасте 3-4 мес. и массой тела 28-30 г. Животных разводили и содержали в условиях вивария Института цитологии и генетики СО РАН (Новосибирск). Воду и стандартный корм (гранулы) животные получали в достаточном количестве. Световой режим был равен 12:12 часам. До начала эксперимента мышей содержали по 6-8 особей в клетках размером 36х23х12 см. За пять дней до эксперимента самцов помещали по одному в одно из двух отделений экспериментальной металлической клетки, размером 28х14х10 см, разделенной пополам прозрачной перегородкой с отверстиями. Предварительно в стенке клетки на расстоянии 3-х см от перегородки было сделано отверстие по размеру диаметра луча лазера (≈9 мм) на уровне тела животного (2,5 см от пола клетки) (Схема 1).

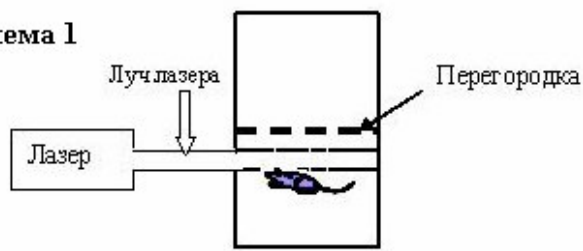

Предполагалось, что волны, пройдя сквозь отверстие, отражались от противоположной стенки клетки, хотя и теряли свою мощность. В день эксперимента животных привозили в помещение, где находился лазер, и после адаптации к новым условиям (не менее одного часа) экспериментальную клетку с мышью подставляли отверстием к лучу лазера. Затем фиксировали поведенческие параметры: а) число подходов и общее время непосредственного обнюхивания и исследования отверстия и пребывания непосредственно возле него; б) общее время и число подходов к перегородке, когда животные находились возле нее, касались передней частью туловища и носом; в) общее время нахождения в зоне действия излучения лазера; г) число и время разрывания/разбрасывания подстилки клетки в различных ее участках в качестве показателей смещенной активности животного. Фиксировали параметры поведения в первые и вторые пять минут облучения, когда соседний отсек был пуст. На третьи 5 минут в свободный отсек клетки помещали другого самца, вызывавшего у хозяина клетки исследовательскую мотивацию [5]. Тестирование поведения животных проводили два экспериментатора, чтобы избежать субъективного влияния на процессы измерения. В качестве контроля в этих экспериментах была исследована группа животных, которой, как и группе опытных мышей, предлагались идентичные экспериментальные процедуры в тех же условиях, за исключением облучения лазером. Другой группе животных в отверстие клетки светили электрическим фонариком, расположенным на расстоянии 10 см от отверстия в клетке. Мощность свечения ≈ 30-40 мВт.

В следующем эксперименте были исследованы эффекты 30 минутного облучения на поведение животных через сутки после

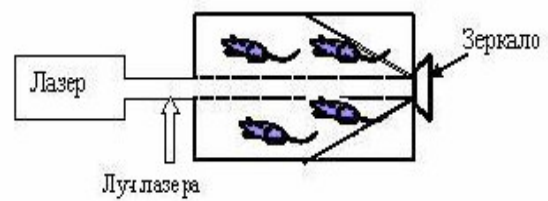

воздействия. Самцов содержали в течение 5 дней по четыре особи в клетке (28х14х10см) с заранее сделанными отверстиями на противоположных стенках клетки на уровне тела животного (2,5 см от пола клетки). В день эксперимента их привозили в домашних клетках в помещение, где находился лазер. При воздействии луч лазера направляли в клетку (Схема 2). Снаружи клетки возле другого отверстия было установлено зеркало, рассеивающее волны внутри клетки. Перемещаясь по клетке, мыши постоянно находились в прямых и отраженных лучах лазера. После облучения мышей возвращали в виварий и рассаживали в индивидуальные клетки. Тестировали животных на следующий день в крестообразном лабиринте, использующемся для оценки состояния тревожности у мышей [6] и представляющем установку с двумя открытыми, двумя закрытыми (огороженными с трех сторон) рукавами, расположенными напротив друг друга, и с центральной площадкой. Лабиринт был приподнят на 1 метр от уровня пола. В качестве параметров поведения регистрировали: число входов/выходов в закрытые и открытые рукава лабиринта и время нахождения в них, а также на центральной площадке. Данные представлялись в процентах от общего времени теста (5 минут) или от общего числа входов/выходов. Были исследованы также некоторые другие формы поведения, обычно регистрируемые в этом тесте (число заглядываний под лабиринт, число переходов из одного закрытого рукава в другой и число выглядываний из закрытого рукава). Все процедуры, за исключением облучения ТГц-волнами, были проделаны и с контрольными животными

Для статистической обработки результатов был использован непараметрический критерий Манна-Уитни, позволяющий сравнивать выборки данных при отсутствии нормального распределения. В каждой группе было 10-12 животных.

## РЕЗУЛЬТАТЫ ИССЛЕДОВАНИЯ

Результаты экспериментов показали, что число подходов к перегородке за первые пять минут было достоверно ниже у опытных самцов, в клетку которых светили через отверстие лазером, по сравнению с контрольными, у которых воздействие отсутствовало ($U=18.5$; $p<0.03$). В другие периоды измерения число подходов и общее время пребывания возле перегородки не отличались достоверно от таковых значений у контрольных животных (Рис. 1, $p>0.05$). Число входов и общее время пребывания животных в зоне действия луча лазера во все последовательные периоды измерения существенно не различались у опытных животных по сравнению с контрольными ($p>0.05$) (Рис. 1). Однако опытные самцы демонстрировали существенно меньший интерес к отверстию, сквозь которое светили лазером. Общее время исследования и обнюхивания отверстия в первые, вторые и третьи пять минут измерения было ниже ($U=13.5$ $p<0.010$; $U=18.0$ $p<0.027$; $U=19.5$ $p<0.037$ - соответственно), как и число подходов к нему ($U=19.5$ $p<0.037$; $U=16.5$ $p<0.020$; $U=15.0$ $p<0.014$ – соответственно) по сравнению с контрольными самцами, которые почти в три раза чаще подходили к отверстию и почти в пять раз дольше находились возле него, исследуя и обнюхивая отверстие. С одной стороны, изменение поведения животных при включении лазера свидетельствует о том, что животные ощущают излучение, с другой стороны, снижение исследовательского поведения по отношению к новому стимулу – лучу лазера, говорит о негативном влиянии, которое оказывает излучение, вызывая избегание животных, то есть, можно говорить об аверсивных свойствах излучения.

Разбрасывание и разрывание подстилки в домашней клетке является сложным поведенческим параметром, интерпретация которого зависит от контекста ситуации. У интактных животных он может характеризовать поведение строительства гнезда или поиска пищи. При стрессирующем воздействии увеличение демонстрации этой формы поведения свидетельствует о смещенной активности, которая часто проявляется при наличии нескольких мотиваций, часть из которых не может быть реализована в силу определенных обстоятельств. В нашем эксперименте (Рис. 1) увеличение числа, особенно во вторые пять минут измерения ($U=15.0$; $p<0.010$), и времени демонстрации ($U=16.0$; $p<0.014$) этой формы поведения может рассматриваться в качестве смещенной активности, вызываемой аверсивными стимулами при невозможности избежать как самого излучения, так и последствий отраженного воздействия ТГц-волнами.

Таким образом, можно сказать, что эффект воздействия излучения лазера на мышей есть, мыши замечают его влияние, при этом ТГц-лучи оказывают, по-видимому, негативное воздействие, вызывая у мышей избегание. Надо отметить, что помещение нового самца в другой отсек клетки приводило к возникновению сильной исследовательской мотивации как у контрольных, так и опытных животных, что было показано ранее [5] и в этом эксперименте: число подходов к перегородке и общее время пребывания возле нее становятся существенно больше по сравне-

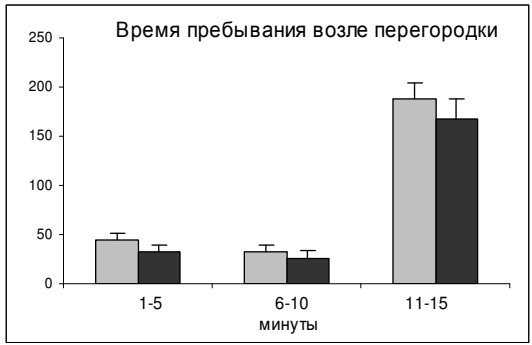 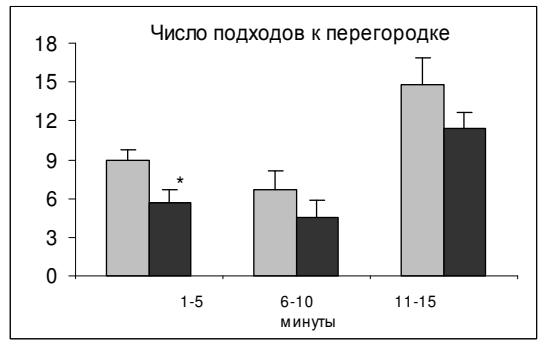
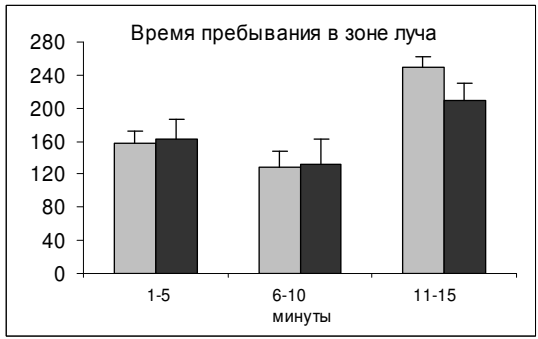 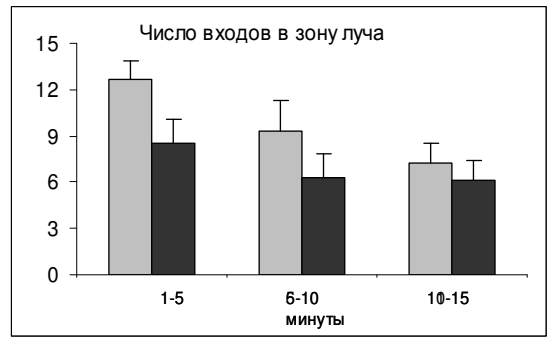
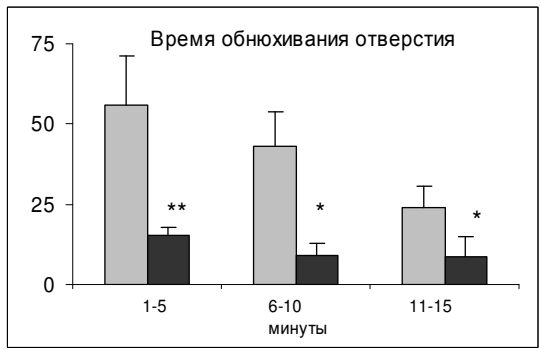 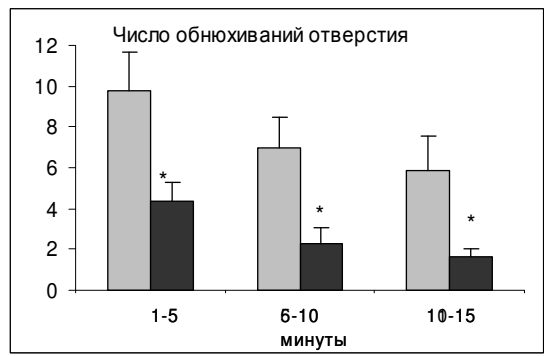
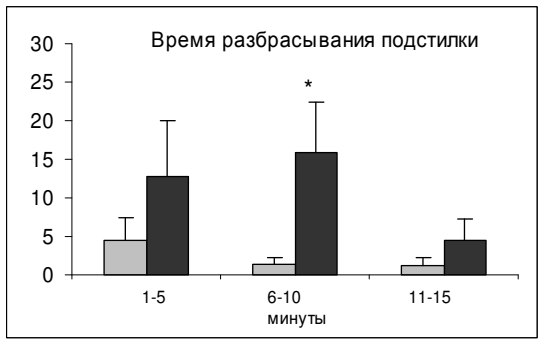 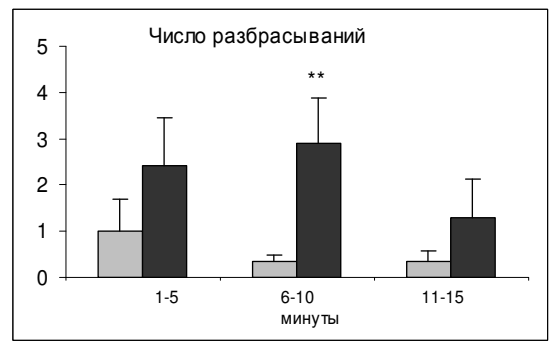

Рисунок 1. Поведение самцов мышей в клетке при действии ТГц лазера (темные столбы – опытные животные) и без него (светлые столбы – контрольные животные), * - *p*<0.05; ** - *p*<0.01, по сравнению с контролем.

Таблица 1. Влияние электрического света на поведение самцов мышей

| Параметры тестов | 1-5 минуты, без партнера | | 6-10 минуты, без партнера | | 11-15 минуты, с партнером | |
|---|---|---|---|---|---|---|
| | Контроль | Свет | Контроль | Свет | Контроль | Свет |
| *Перегородка* | | | | | | |
| число подходов | 5.5±1.4 | 8.5±2.0 | 5.5±1.2 | 7.1±1.1 | 9.5±1.6 | 10.6±1.2 |
| время, сек | 48.5±12.0 | 55.9±12.0 | 43.4±15.5 | 65.6±17.5 | 201.8±16.4 | 205.0±7.0 |
| *Нахождение в зоне действия излучения* | | | | | | |
| число | 7.5±1.2 | 10.9±2.08 | 7.0±1.5 | 9.5±1.6 | 4.6±1.1 | 6.3±1.0 |
| время, сек | 191.0±22.1 | 168.5±14.8 | 163.4±24.0 | 144.6±21.8 | 266.3±7.6 | 260.1±5.8 |
| *Обнюхивание и исследование отверстия с лучом света* | | | | | | |
| число | 5.0±1.2 | 6.0±1.4 | 4.2±1.1 | 4.9±0.8 | 3.8±1.1 | 2.6±0.6 |
| время, сек | 24.8±8.0 | 28.5±6.7 | 16.5±5.9 | 18.4±5.1 | 9.6±2.6 | 8.1±2.9 |
| *Разбрасывание подстилки* | | | | | | |
| число | 1.0±0.7 | 0.3±0.2 | 1.2±0.9 | 0.6±0.5 | 1.3±0.8 | 0.1±0.1 |
| время, сек | 5.5±5.1 | 2.0±1.3 | 3.7±3.3 | 0.5±0.4 | 3.8±2.4 | 0.4±0.4 |

нию с этими параметрами, когда соседний отсек был пуст (Рис. 1, $p<0.01$). Однако и при наличии другой мотивации, время исследования выходного отверстия лазера было по-прежнему ниже у опытных животных по сравнению с контрольными.

В аналогичных условиях электрический свет от фонарика не оказал достоверного влияния на поведение животных, у которых все поведенческие параметры мало отличались у опытных и контрольных самцов (Таблица 1). Таким образом, влияние света было менее значительно и не вызывало избегания.

Отставленный эффект свечения лазером в течение 30 минут (Схема 2) сказался на некоторых параметрах поведения, исследованных на следующий день в крестообразном лабиринте (Таблица 2). У опытных самцов число выходов в центр было ниже ($U=25.0$; $p<0.02$), а число заходов в закрытые рукава выше ($U=23.5$; $p<0.01$), чем у контрольных животных, не подвергавшихся облучению лазером, что свидетельствует о большей тревожности опытных животных. Они вообще не выходили в открытые рукава. Различия по числу заходов и времени нахождения в открытых рукавах между

Таблица 2. Эффект отставленного действия облучения на поведение самцов мышей в тесте крестообразного лабиринта.

| Параметры поведения,% | Контроль | После облучения |
|---|---|---|
| Открытые рукава, n | 2.9 ± 1.7 | 0.0 ± 0.0+ |
| Открытые рукава, t | 0.51 ± 0.3 | 0.0 ± 0.0+ |
| Центр, n | 49.2 ± 0.7 | 46.4 ± 1.0 * |
| Центр, t | 12.1 ± 1.5 | 9.5 ± 1.3 |
| Закрытые рукава, n | 47.8 ± 2.9 | 53.3 ±0.9 ** |
| Закрытые рукава, t | 87.4 ± 1.7 | 89.9 ± 1.4 |

Приведены производные показатели числа (n) и времени (t) поведенческих параметров, выраженные в процентах от общего числа входов/выходов в рукава и времени теста (5 минут), соответственно. * - $p<0.05$; ** - $p<0.01$, + $p < 0.08$, vs контроль.

опытными и контрольными самцами были на уровне тенденции ($U=40.0$; $p<0.08$). Число выглядываний из закрытого рукава, число заглядываний под лабиринт и число переходов из одного закрытого рукава в другой у опытной и контрольной групп существенно не отличались ($p>0.05$)

Подводя итог, можно сказать, что ТГц-лучи даже при кратковременном действии оказывают влияние на мышей, которые чувствуют воздействие излучения. Это позволяет использовать некоторые поведенческие параметры, в частности, избегание местонахождения возле луча, в качестве биодетекторов воздействия ТГц-лучей на мышей. Каковы механизмы и на каком уровне физиологического реагирования терагерцевое излучение оказывает эффект на организм, предположить в настоящий момент не представляется возможным. Поскольку наиболее чувствительным параметром оказалось число подходов и время исследования (обнюхивания) отверстия с лучом лазера, можно думать, что наибольший эффект воздействия ТГц-волнами был при непосредственном контакте с лазером.

Об этом же свидетельствует отсутствие существенного влияния излучения на общее время пребывания в зоне действия лучей лазера, которое у опытных и контрольных животных был сходным. Можно полагать, что воздействие ТГц-волн в меньшей степени проявляется на расстоянии, и действует непосредственно на ткани, клетки и биологические молекулы, что и было показано ранее в некоторых экспериментах [4, 8, 9]. Облучение в течение 30 минут сказалось на эмоциональном состоянии животных. По сравнению с контрольными животными у облученных мышей был несколько выше уровень тревожности, оцениваемый на следующий день после воздействия.